\documentclass[10pt]{article}
\usepackage{graphicx}
\usepackage{amsmath}
\usepackage{enumerate}
\usepackage{natbib}
\usepackage{url}
\usepackage{amsfonts}
\usepackage{algpseudocode}
\usepackage{algorithm}

\newcommand{\blind}{0}

\begin{document}

\bibliographystyle{agsm}

\def\spacingset#1{\renewcommand{\baselinestretch}%
{#1}\small\normalsize} \spacingset{1}


\if0\blind
{
  \title{\bf Unimodal clustering using isotonic regression: ISO-SPLIT}
  \author{Jeremy F. Magland\hspace{.2cm}\\
    Simons Center for Data Analysis\\
    and \\
    Alex H. Barnett \\
    Simons Center for Data Analysis \\ and Department of Mathematics, Dartmouth College}
  \maketitle
} \fi

\if1\blind
{
  \bigskip
  \bigskip
  \bigskip
  \begin{center}
    {\LARGE\bf Unimodal clustering using isotonic regression: ISO-SPLIT}
  \end{center}
  \medskip
} \fi

\bigskip
\begin{abstract}
A limitation of many clustering algorithms is the requirement to tune adjustable parameters for each application or even for each dataset. Some techniques require an \emph{a priori} estimate of the number of clusters while density-based techniques usually require a scale parameter. Other parametric methods, such as mixture modeling, make assumptions about the underlying cluster distributions. Here we introduce a non-parametric clustering method that does not involve tunable parameters and only assumes that clusters are unimodal, in the sense that they have a single point of maximal density when projected onto any line, and that clusters are separated from one another by a separating hyperplane of relatively lower density. The technique uses a non-parametric variant of Hartigan's dip statistic using isotonic regression as the kernel operation repeated at every iteration. We compare the method against k-means++, DBSCAN, and Gaussian mixture methods and show in simulations that it performs better than these standard methods in many situations. The algorithm is suited for low-dimensional datasets with a large number of observations, and was motivated by the problem of ``spike sorting'' in neural electrical recordings. Source code is freely available.
\end{abstract}

\noindent%
\vfill

\newpage
\spacingset{1.45} 

\section {Introduction}

Unsupervised data clustering is a methodology for automatically partitioning a set of data points in a manner that reflects the underlying structure of the data. In many clustering applications with continuous data in $p$ dimensions, clusters are expected to have a core region of high density and to be separated from one another by a region of relatively lower density. The motivating application for the authors is spike sorting of neuron firing events recorded electrically, for which this property has been found to hold experimentally \citep{tiganj,vargas}. We expect this work to be useful in a variety of other clustering applications, particularly in cases with a large number of observations in a low dimensional feature space.

A limitation of most clustering algorithms is the need to tune a set of adjustable parameters. The adjustments may be per application, or even per dataset. For k-means \citep{kmeans}, the adjustable parameter is $K$, the prospectively estimated number of clusters. For large datasets where dozens of clusters are present, the choice of $K$ is especially problematic. In addition, the output of k-means depends heavily on the initialization step (choosing seed points), and the algorithm is often repeated several times to obtain a more globally optimal solution. K-means++ \citep{kmeanspp} does a better job at seeding, but some rerunning is still required. Even with optimal seeding, if some clusters are small or sparse relative to the dominant cluster then they are often merged into nearby clusters. In general, k-means tends to favor artificially splitting larger clusters at the expense of merging smaller ones. A further limitation is that the algorithm assumes isotropic cluster distributions with equal populations and equal variances.

Gaussian mixture modeling (GMM), usually solved using expectation-maximization (EM) iterations \citep{em}, is more flexible than k-means since it allows each cluster to be assigned its own multivariate normal distribution. Many variations exist, some of which are outlined in \citep[Ch.~11]{murphy}. While some implementations require prospective knowledge of the number of clusters \citep[Ch.~8]{mixturemodels}, other implementations consider this as a free variable \citep[e.g.][]{roberts1998bayesian}. 
The main limitation is that clusters must be well modeled by Gaussians; furthermore, as in k-means, it can be difficult to find the global solution, especially when the number of clusters is large.
Recently, mixture models with skew non-Gaussian
components have been developed \citep[e.g.][]{skewGMM,skewGMM2}. However, the increase in model complexity
results in a more challenging optimization problem.

Hierarchical clustering \citep[Ch.~14]{zaki-book} does not require specification of the number of clusters ahead of time, but this is because the output is a dendrogram rather than a partition. Thus it cannot immediately be applied to our application of interest. There is a way to obtain an automated partition from the dendrogram, but this involves specifying a criteria for cutting the binary tree (much like specifying $K$). Other choices need to be made for agglomerative hierarchical clustering in order to determine which clusters are merged at each iteration. Furthermore, hierarchical clustering has time complexity at least $O(n^2)$ where $n$ is the size of the dataset \citep[Sec.~14.2.3]{zaki-book}.

Density-based clustering techniques such as DBSCAN \citep{dbscan} are promising since they do not make assumptions about
data distributions, 
so they can handle clusters with arbitrary non-convex shapes. The drawback is that two parameters must be adjusted depending on the application, including $\epsilon$, a parameter of scale. The algorithm is especially sensitive to this parameter in higher dimensions. A further limitation is that if the clusters
substantially differ in density, then no choice of $\epsilon$ will simultaneously handle the entire dataset. Thus a scale-independent method using data density is desirable.

Other density-based techniques, such as \citep{mean-shift}, involve the initial step of constructing a continuous non-parametric probability density function \citep[Ch.~15]{zaki-book}. The basic version of the kernel density method \citep{kernel-density-function-1,kernel-density-function-2} involves specifying a spatial scale parameter (the so-called bandwidth), and suffers from the same problem as DBSCAN. Variations of this method can automatically estimate an optimal bandwidth \citep{silverman-density-estimation}, and can even derive a value that is spatially dependent. There are many density-estimation methods to choose from \citep[incluuding][]{rodriguez-clustering}, but they often depend on adjustable distance parameters.
In general, these methods become computationally intractable in higher dimensions.

Here we introduce an efficient density-based, scale-independent clustering technique suited for situations where clusters are expected to be unimodal and when any pair of distinct clusters may be separated by a hyperplane. We say a cluster is {\em unimodal} if it arises from a distribution that has a single point of maximum density when projected onto ony line. Thus, our assumption is that the projection of any two clusters onto the normal of the dividing hyperplane gives a bimodal distribution separating the clusters at its local minimum. Loosely speaking, this is guaranteed when the clusters are sufficiently spaced and have convex shapes.

In addition to being density-based, our technique has the flavor of agglomerative hierarchical clustering as well as the EM-style iterative approach of k-means. The algorithm uses a non-parametric procedure for splitting one-dimensional distributions based on a modified Hartigan's dip statistic \citep{hartigan1985dip} and isotonic regression. Neither isotonic regression nor Hartigan's method involve adjustable parameters (aside from selection of a statistical significance threshold). In particular, no scale parameter is needed for density estimation. Furthermore, since the core step performed at each iteration is one-dimensional (1D) clustering applied to projections of data subsets onto lines, we avoid the curse of dimensionality (the tradeoff being that we cannot handle clusters of arbitrary shape).

This paper is organized as follows. First we describe an algorithm for splitting a 1D sample into unimodal clusters. This procedure forms the basis of the $p$-dimensional clustering technique, ISO-SPLIT, defined in Section 3. Simulation results are presented in Section 4, comparing ISO-SPLIT with three standard clustering techniques. In addition to quantitative comparisons using a measure of accuracy, examples illustrate situations where each algorithm performs best. The fifth section is an application of the algorithm to spike sorting of neuronal data. Next we discuss computational efficiency and scaling properties. Finally, Section 8 summarizes the results and discusses the limitations of the method. The appendices cover implementation details for isotonic regression, generation of synthetic datasets for simulations, and provide evidence for insensitivity to parameter adjustments.

\begin{figure}
\begin{center}
\includegraphics[width=5.5in]{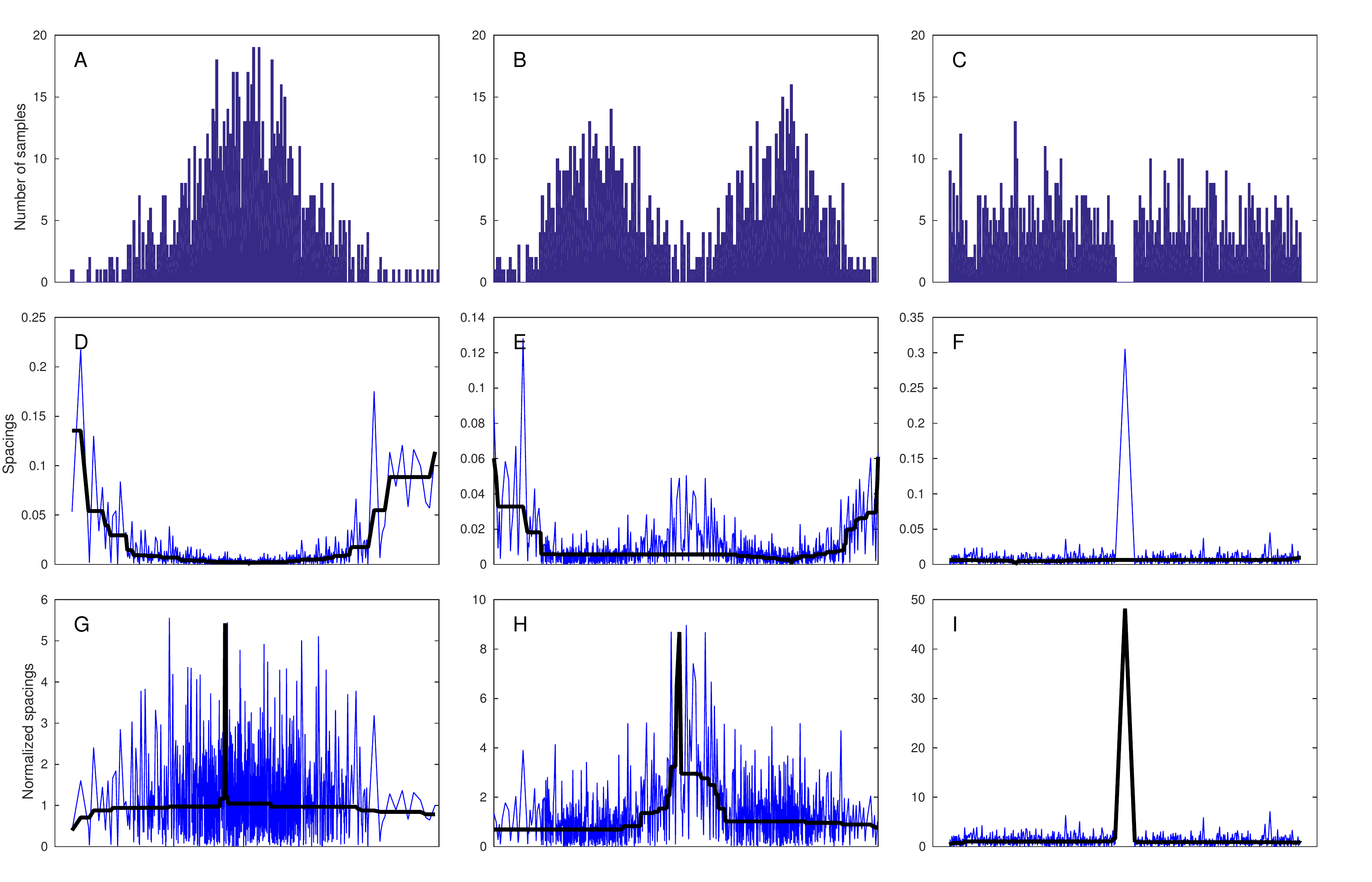}
\end{center}
\caption{
(A,B,C) Histograms for three simulated 1D distributions with $n=1000$ datapoints. The first represents a single cluster whereas the second and third are sampled from bimodal probability distributions. (D,E,F) show the corresponding sequence of spacings (finite differences) between adjacent points. The dark curves are the outputs of down-up isotonic regression (defined in Section \ref{clustering_1d} and Appendix \ref{appendixUpdown}) for approximating these series. (G,H,I) show the corresponding normalized spacings that are fit using up-down isotonic regression. The peaks in the dark curves of (H) and (I) indicate that the set of samples should be split into two clusters, whereas the dip in (G) is not used to split distribution since the unimodality hypothesis in this case is not rejected.
}
\label{fig:plots_1d}
\end{figure}

\section {Clustering in one dimension}
\label{clustering_1d}

Any approach overcoming the above limitations must at least be able to do so in the simplest, 1D case ($p=1$). Here we develop a non-parametric approach to 1D clustering using a statistical test for unimodality and isotonic regression. The procedure will then be used as the kernel operation in the more general situation ($p\geq2$) as described in Section~\ref{isosplit-algorithm}.

Clustering in 1D is unique because the input data can be sorted. The problem reduces to selecting cut points (real numbers between adjacent data points) that split the data into $K$ clusters, each corresponding to an interval on the real line. We assume that the clusters are unimodal so that adjacent clusters are separated by a region of relatively lower density. For simplicity we will describe an algorithm for deciding whether there is one cluster, or more than one cluster. In the latter case, a single cut point is determined representing the boundary separating one pair of adjacent clusters. Note that once the data have been split, the same algorithm may then be applied recursively on the left and right portions leading to further subdivisions, converging when no more splitting occurs. Thus the algorithm described here may be used as a basis for general 1D clustering.

Let $x_1<\dots<x_n$ be the sorted (assumed distinct\footnote{The data are assumed to be independent samples from a continuous probability distribution, thus distinct with probability one. More details are given in the discussion.}) real numbers (input data samples). Our fundamental assumption is that two adjacent clusters are always separated by a region of lower density. In other words, if $a_1$ and $a_2$ are the \emph{centers} of two adjacent 1D clusters, then there exists a cut point $a_1<c<a_2$ such that the density near $c$ is significantly less than the densities near both $a_1$ and $a_2$. The challenge is to define the notion of density near a point. The usual approach is to use histogram binning or kernel density methods. However, as described above, we want to avoid choosing a length scale $\epsilon$.

Instead we use a variant of Hartigan's dip test. The null hypothesis is that the set $X=\{x_j\}$ is an independent sampling of a unimodal probability density $f(x)$, which by definition is increasing on $[-\infty,c]$ and decreasing on $[c,\infty]$. The dip statistic is the Kolmogorov-Smirnov distance
$$D_X = \{\sup_x |F(x)-S_X(x)|\},$$
where
$$S_X(x)=\#\{j:x_j\leq x\}$$
is the empirical distribution and $F$ is a unimodal cumulative distribution function that approximates $S_X$. In contrast to Hartigan's original method, we use down-up isotonic regression to determine $F$. As outlined in Appendix \ref{appendixUpdown}, down-up isotonic regression finds the best least-squares approximation to a function that is monotonically decreasing to a critical point $c$ and then monotonically increasing. This algorithm is used to fit the sequence of spacings between adjacent points, as illustrated in Figure~\ref{fig:plots_1d}. The spacings attain their minimum at the peak of the distribution. The function is then integrated to obtain the cumulative distribution function $F$. If the $D_X$ statistic lies above a certain threshold $\tau_{n}$, then the unimodality hypothesis is rejected. In this study we used a threshold of $\tau_{n}=\alpha/\sqrt{n}$ with $\alpha=1.2$ (see Appendix \ref{appendixSensitivity}).

Hartigan's test only produces an accept/reject result, and does not supply an optimal cut point separating the clusters. Let $x_1<\dots<x_n$ be the sorted (assumed distinct) real numbers (input data samples). Our goal is to obtain such a cut point $c$. Let $s_j$ denote the spacing between adjacent points $x_j$ and $x_{j+1}$, which is an approximation to the reciprocal of density at this location (we assume that the $x_j$ are all distinct). To detect a density dip, we normalize these spacings by setting 
$$\tilde{s}_j=\frac{s_j}{t_j},$$
where $t_j$ is the corresponding spacing for the approximating unimodal distribution obtained above. As shown in \ref{fig:plots_1d}, we can then fit the new sequence of spacings using up-down isotonic regression, since we expect the normalized spacings to increase at the dip. The cut point is selected at the peak of this up-down fit to the spacings. 

There is a flaw with Hartigan's test in the case where the number of points in one cluster (say on the far left) is very small compared with the total size $n$. This is because the absolute size of the dip depends only on the region at the interface between the two clusters, whereas the test for rejection becomes increasingly stringent with increasing $n$. To remedy this we compute a series of dip tests of sizes $4,8,16,32,\dots,\lfloor\log_2 n\rfloor$. For each size, two tests are performed, one starting from the left (more negative) side of the dataset, and one starting from the right (more positive). As soon as the unimodality hypothesis is rejected in one of these tests, the algorithm halts and the null hypothesis is rejected, and the cut point for that segment is returned. Otherwise it is accepted.

\begin{figure}
\begin{center}
\includegraphics[width=5.5in]{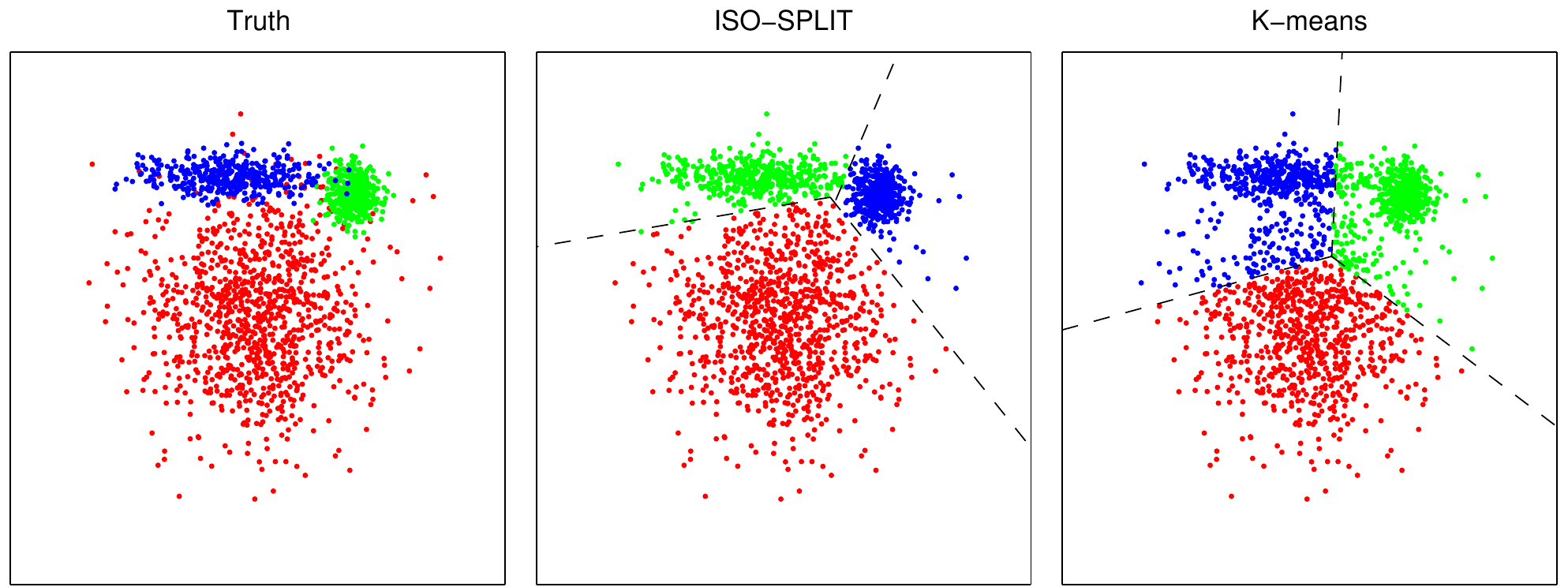}
\end{center}
\caption{
Illustration that the
placement of decision boundaries for ISO-SPLIT is more flexible than for k-means clustering.
The left panel shows an artificial 2D dataset where each of the three labels has points drawn from bivariate Gaussian pdfs. The middle panel shows the result of clustering with ISO-SPLIT, and the decision lines (in general hyperplanes).
The right panel shows the same using k-means. Unlike in k-means, which assumes all clusters have the same variance, in ISO-SPLIT hyperplanes may be positioned closer to one of the centroids than the other, and
need not be orthogonal to the line connecting centroids. (See color online.)
}
\label{fig:decision_boundaries}
\end{figure}

\section {Clustering in higher dimensions using one-dimensional projections}

\label{isosplit-algorithm}

In this section we address the $p$-dimensional situation ($p\geq2$) and describe an iterative procedure, termed ISO-SPLIT, in which the 1D routine is repeated as a kernel operation. The decision boundaries are less restrictive than $k$-means which always splits space into Voronoi cells with respect to the centroids, as illustrated in Figure~\ref{fig:decision_boundaries}.

The proposed procedure is outlined in Algorithm~\ref{alg:main_algorithm}. The input is a collection of $n$ points in $\mathbb{R}^p$, and the output is the collection of corresponding labels (or cluster memberships). The approach is similar to agglomerative hierarchical methods in that we start with a large number of clusters (output of \emph{InitializeLabels}) and iteratively reduce the number of clusters until convergence. However, in addition to merging clusters the algorithm may also redistribute data points between adjacent clusters. This is in contrast to agglomerative hierarchical methods. At each iteration, the two \emph{closest} clusters (that have not yet been handled) are selected and all data points from the two sets are projected onto a line orthogonal to the proposed hyperplane of separation. The 1D split test from the previous section is applied (see above) and then the points are redistributed based on the optimal cut point, or if no statistically significant cut point is found the clusters are merged. This procedure is repeated until all pairs of clusters are handled.

\spacingset{0.95} 
\algrenewcomment[1]{\(\triangleright\) #1}
\begin{algorithm}
\caption{}
\begin{algorithmic}
\Function{ISO-SPLIT}{$\{y_1,\dots,y_n\}$,$\alpha$}
\State $\{L_1,\dots,L_n\} \gets \text{InitializeLabels} (\{y_1,\dots,y_n\})$
\State $\text{UsedPairs} \gets \{\}$
\Loop
	\State $[k_1,k_2,\text{exists}] \gets \text{FindClosestPair}(\{y_1,\dots,y_n\},\{L_1,\dots,L_n\},\text{UsedPairs})$
	\If{$\text{not }\text{exists}$} \Comment{Converged}
		\State $\text{\textbf{break}}$ 
	\EndIf
	\State $S_1=\{j: L_j=k_1\}$
	\State $S_2=\{j: L_j=k_2\}$
	\State $V \gets \text{GetProjectionDirection}(S_1,S_2)$
	\State $X_1 \gets \{\text{Project}(V,y_j): j\in S_1\}$
	\State $X_2 \gets \{\text{Project}(V,y_j): j\in S_2\}$
	\State $[c,\text{do\_split}] \gets \text{ComputeOptimalCutpoint}(X_1\cup X_2)$
	\If{$\text{do\_split}$} \Comment{Redistribute according to cut point}
		\ForAll{$j \in S_1\cup S_2$} 
			\If{$\text{Project}(L,y_j)\leq c$}
				$L_j \gets k_1$
			\Else
				$\text{ }L_j \gets k_2$
			\EndIf
		\EndFor
	\Else \Comment{ Merge}
		\State $L_j\gets k_1 \text{ }\forall j\in S_1\cup S_2$ 
	\EndIf
	\State $\text{UsedPairs}\gets\text{UsedPairs}\cup\{(S_1,S_2),(S_2,S_1)\}$
        \EndLoop
\State\Comment{Reassign labels to be in $\{1,\dots,K\}$}
\State $L\gets \text{Remap}(L)$
\State \Return $L$
\EndFunction
\end{algorithmic}
\label{alg:main_algorithm}
\end{algorithm}
\spacingset{1.45} 

The best line of projection may be chosen in various ways. The simplest approach is to use the line connecting the centroids of the two clusters of interest. Although this choice may be sufficient in most situations, the optimal hyperplane of separation may not be orthogonal to this line. The approach we used in our implementation is to estimate the covariance matrix of the data in the two clusters (assuming Gaussian distributions with equal variances) and use this to whiten the data prior to using the above method. The function \emph{GetProjectionDirection} in Algorithm~\ref{alg:main_algorithm} returns a unit vector $V$ representing the direction of the optimal projection line, and the function \emph{Project} simply returns the inner product of this vector with each data point.

Similarly, there are various approaches for choosing the closest pair of clusters at each iteration (\emph{FindClosestPair}). One way is to minimize the distance between the two cluster centroids. Note, however, that we don't want to repeat the same 1D kernel operation more than once. Therefore, the closest pair that has not yet been handled is chosen. In order to avoid excessive iterations we used a heuristic for determining whether a particular cluster pair (or something very close to it) had been previously attempted.

The function \emph{InitializeLabels} creates an initial labeling (or partition) of the data. This may be implemented using the $k$-means algorithm with the number of initial clusters chosen to be much larger than the expected number of clusters in the dataset, the assumption being that the output should not be sensitive once $K_\text{initial}$ is large enough (see Appendix \ref{appendixSensitivity}). For our tests we used the minimum of $20$ and four times the true number of clusters. Since datasets may always be constructed such that our choice of $K_\text{initial}$ is not large enough, we will seek to improve this initialization step in future work.

The critical step is \emph{ComputeOptimalCutpoint}, which is the 1D clustering procedure described in the previous section, using a threshold of $\tau_n=\alpha/\sqrt{n}$.


\begin{figure}
\begin{center}
\includegraphics[width=5.5in]{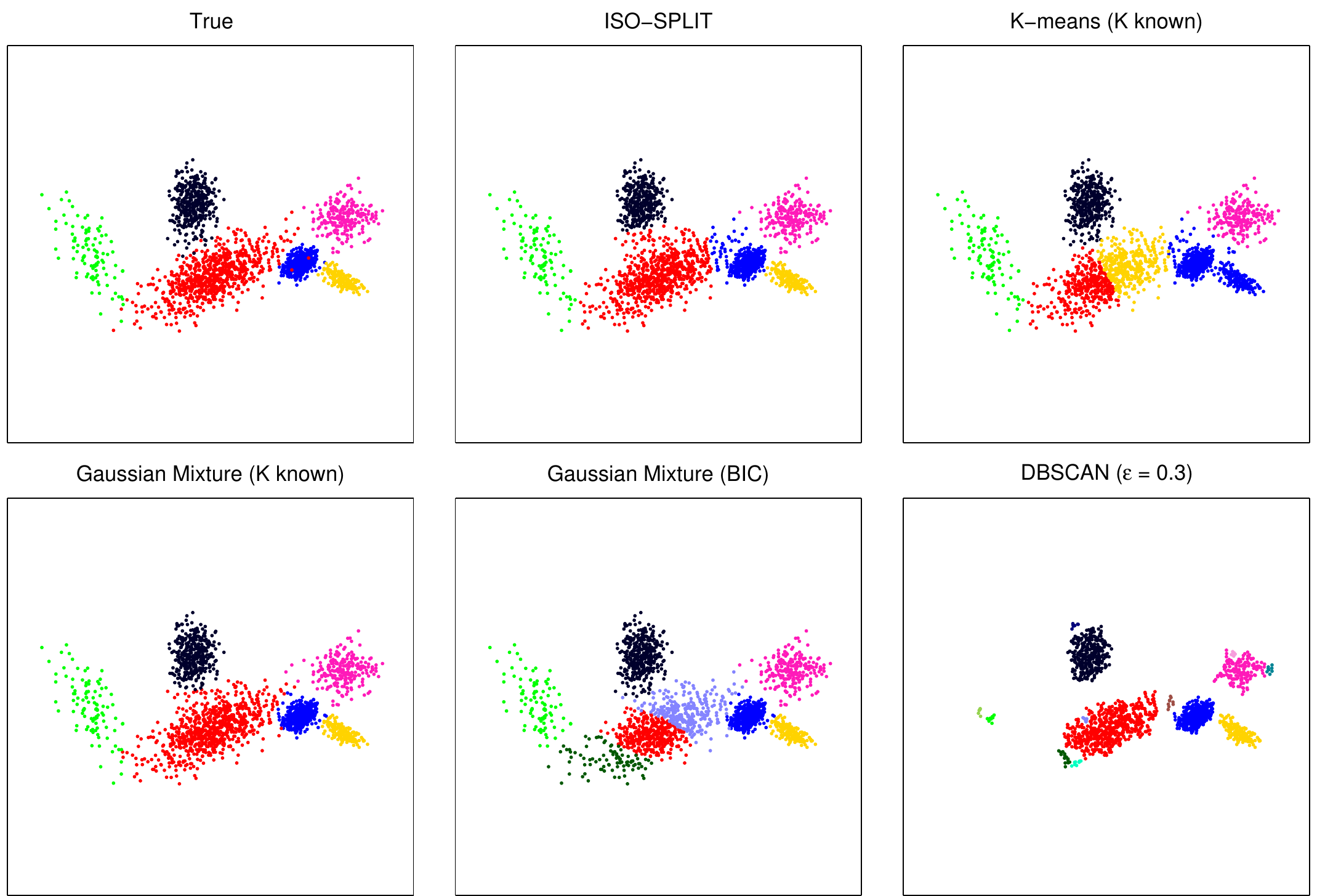}
\end{center}
\caption{
K-means assumes that the cluster populations and variances are all the same. In this 2D example (corresponding to the $K=6$ case of \textbf{Simulation 2 (Anisotropic)} below), two relatively small clusters are merged by k-means while a larger cluster is split. On the other hand, ISO-SPLIT makes no assumptions about relative cluster sizes, thus is able to handle this situation. Gaussian mixture clustering with unknown $K$ (determined using BIC) also fails in this case. DBSCAN fails because the clusters do not all have the same density. Note that ISO-SPLIT was not given information about the true number of clusters nor the expected density.
}
\label{fig:simulation2}
\end{figure}

\begin{figure}
\begin{center}
\includegraphics[width=5.5in]{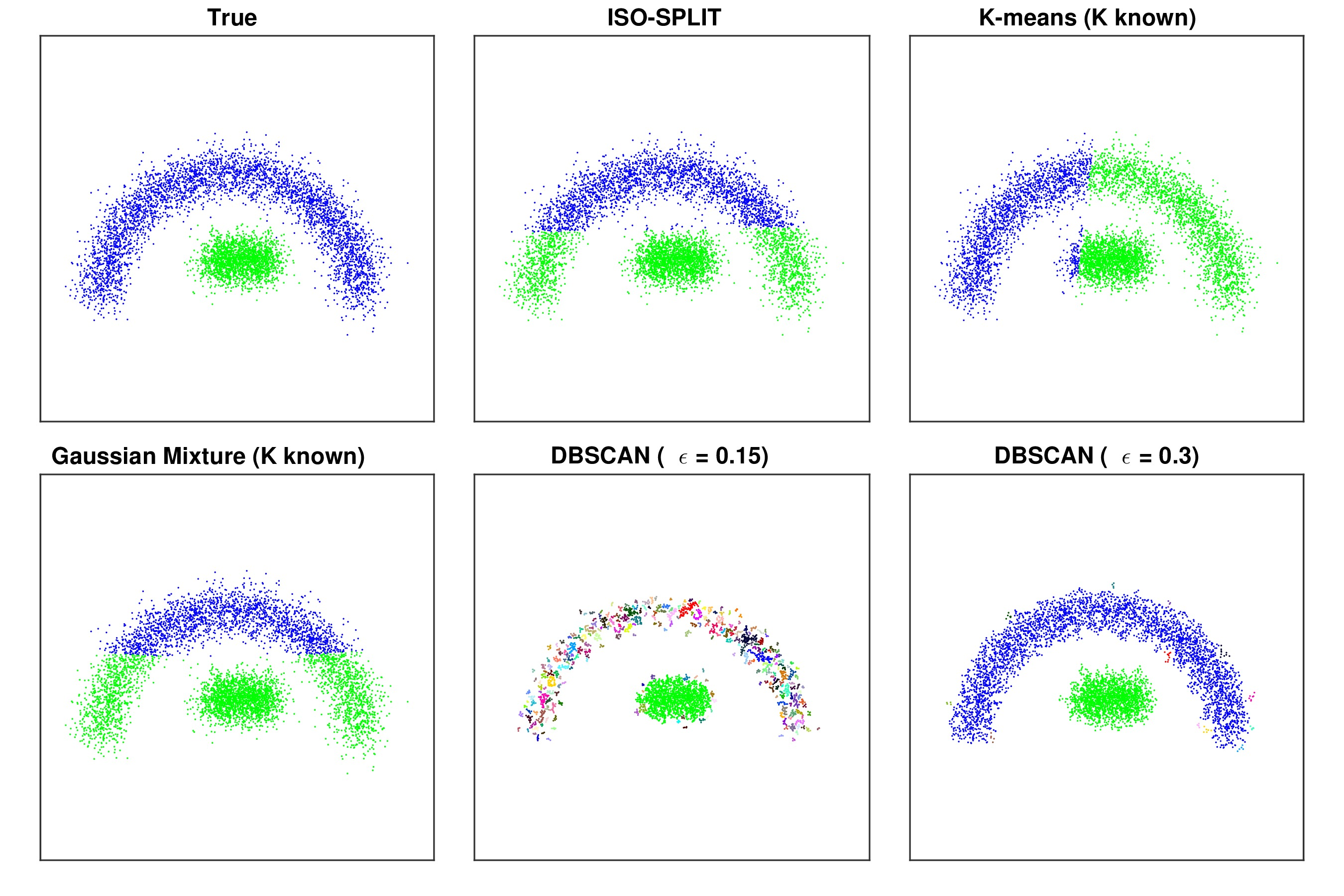}
\end{center}
\caption{
Clustering of synthetic 2D data with $K=2$ clusters (one of which is non-convex) which cannot be separated by a hyperplane.
Unlike ISO-SPLIT, DBSCAN can handle this situation well. However, as the last two panels show, the scale parameter $\epsilon$ must be properly chosen.
}
\label{fig:example_dbscan}
\end{figure}

Figure \ref{fig:simulation2} highlights a case where ISO-SPLIT outperforms k-means, DBSCAN, and GMM (for unknown $K$). This example was selected from Simulation 2 as will be described in Section~\ref{algorithm_comparison}. Unlike k-means and DBSCAN, ISO-SPLIT makes no assumptions about relative cluster population sizes and peak densities. On the other hand, Figure \ref{fig:example_dbscan} illustrates a case where DBSCAN (when properly tuned) performs better than the other methods.

\begin{figure}
\begin{center}
\includegraphics[width=2.5in]{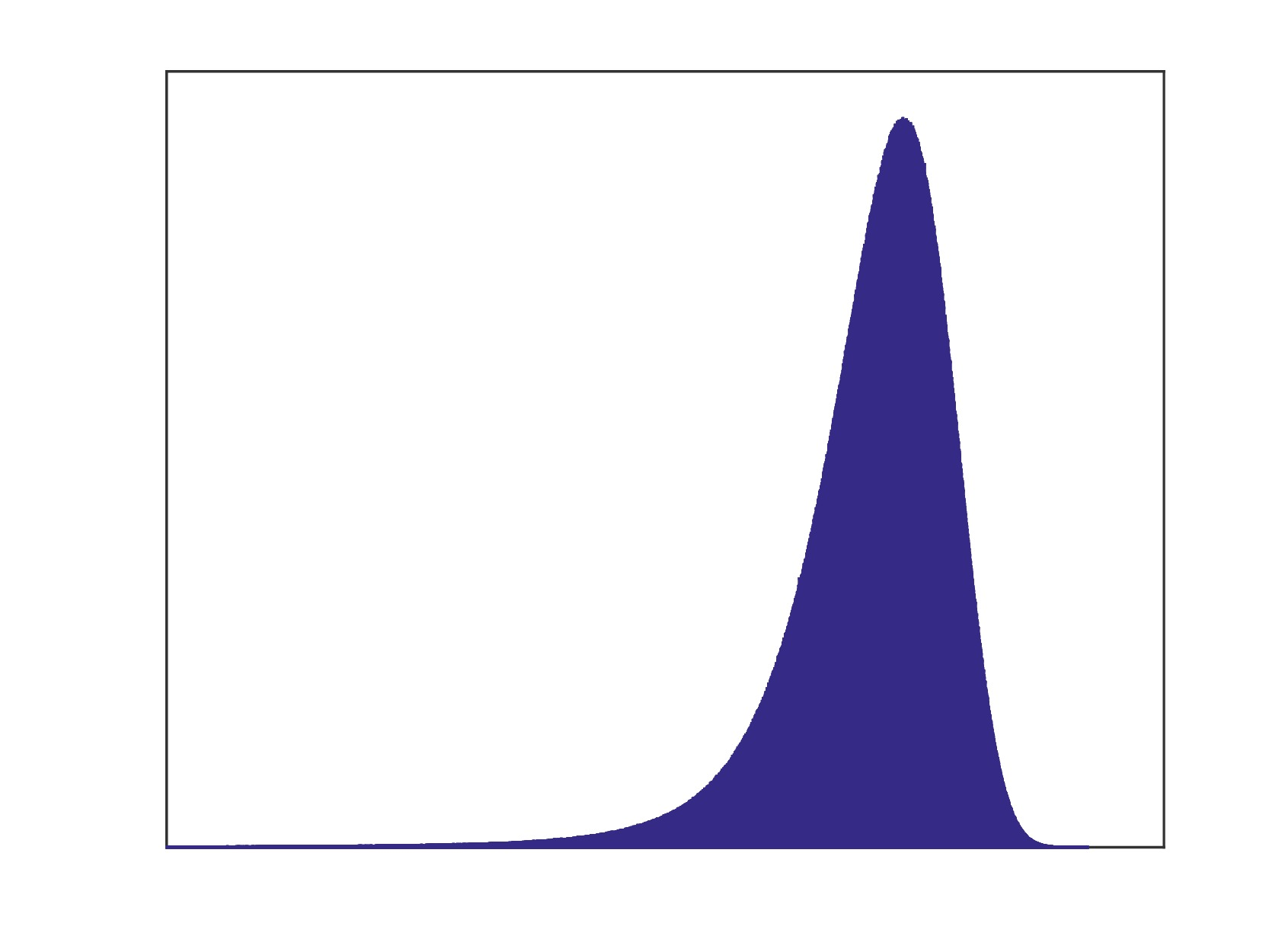}
\end{center}
\caption{
Histogram of the 1D probability distribution used to generate skewed, non-Gaussian clusters in Simulation 3. See Equation \eqref{eq:nongaussian}.
}
\label{fig:nongaussian_histogram}
\end{figure}

\begin{figure}
\begin{center}
\includegraphics[width=5.5in]{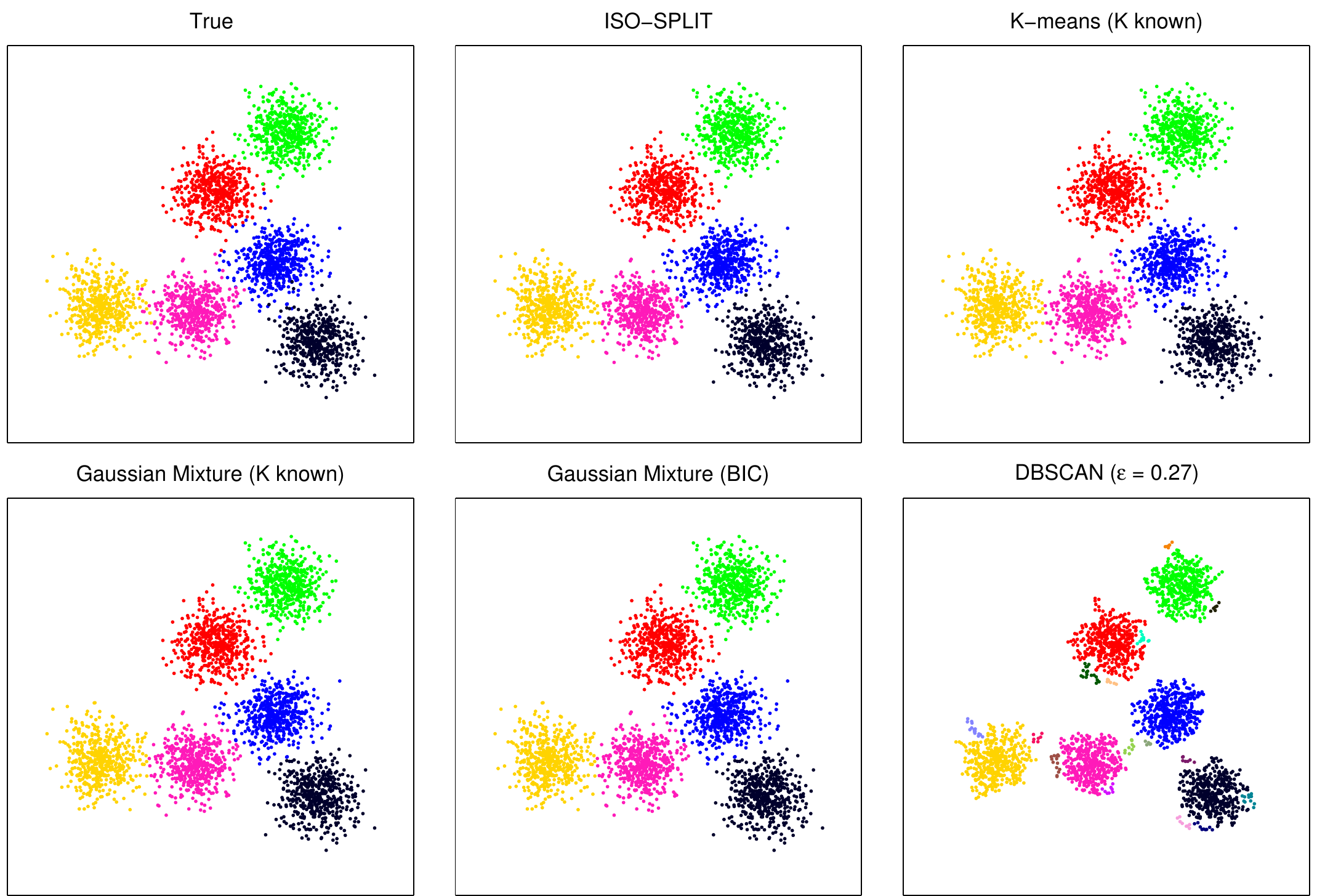}
\end{center}
\caption{
Example run from \textbf{Simulation 1 (Isotropic)}. All investigated approaches perform well when clusters are well-spaced and isotropic with equal variances and populations.
}
\label{fig:simulation1}
\end{figure}

\begin{figure}
\begin{center}
\includegraphics[width=5.5in]{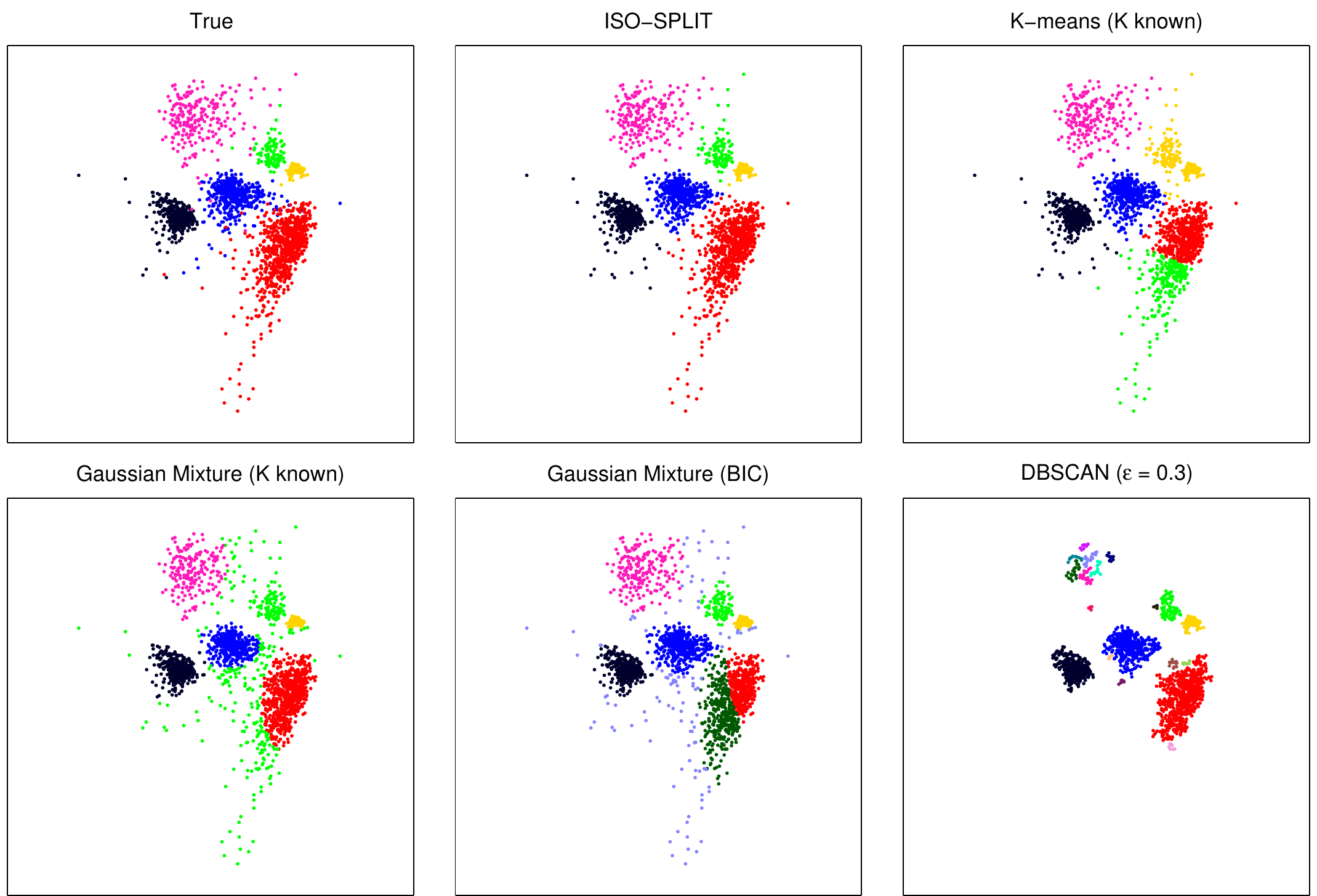}
\end{center}
\caption{
Example run from \textbf{Simulation 3}, with non-Gaussian (skewed) clusters. ISO-SPLIT is density-based and able to handle this situation, whereas k-means and GMM fail due to violation of their underlying assumptions. Despite being density based, DBSCAN also fails since it cannot handle datasets containing clusters of widely differing densities.
}
\label{fig:simulation3}
\end{figure}

\begin{figure}
\begin{center}
\includegraphics[width=5.5in]{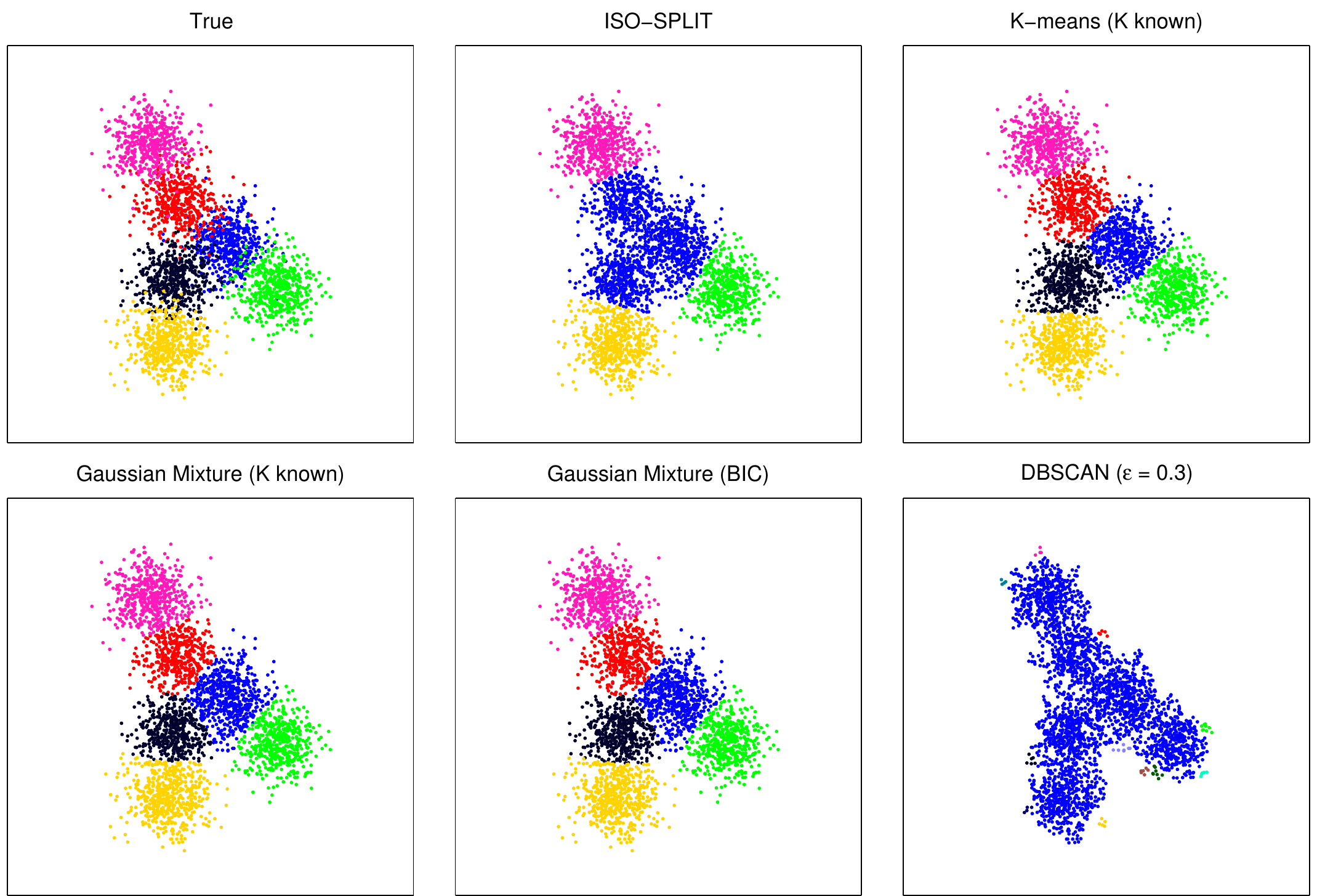}
\end{center}
\caption{
Example run from \textbf{Simulation 4} in which clusters are tightly packed ($z_0=1.7$). The two density based methods (ISO-SPLIT and DBSCAN) erroneously merge adjacent clusters in this case because they are not sufficiently separated by density.
}
\label{fig:simulation4}
\end{figure}

\begin{figure}
\begin{center}
\includegraphics[width=5.5in]{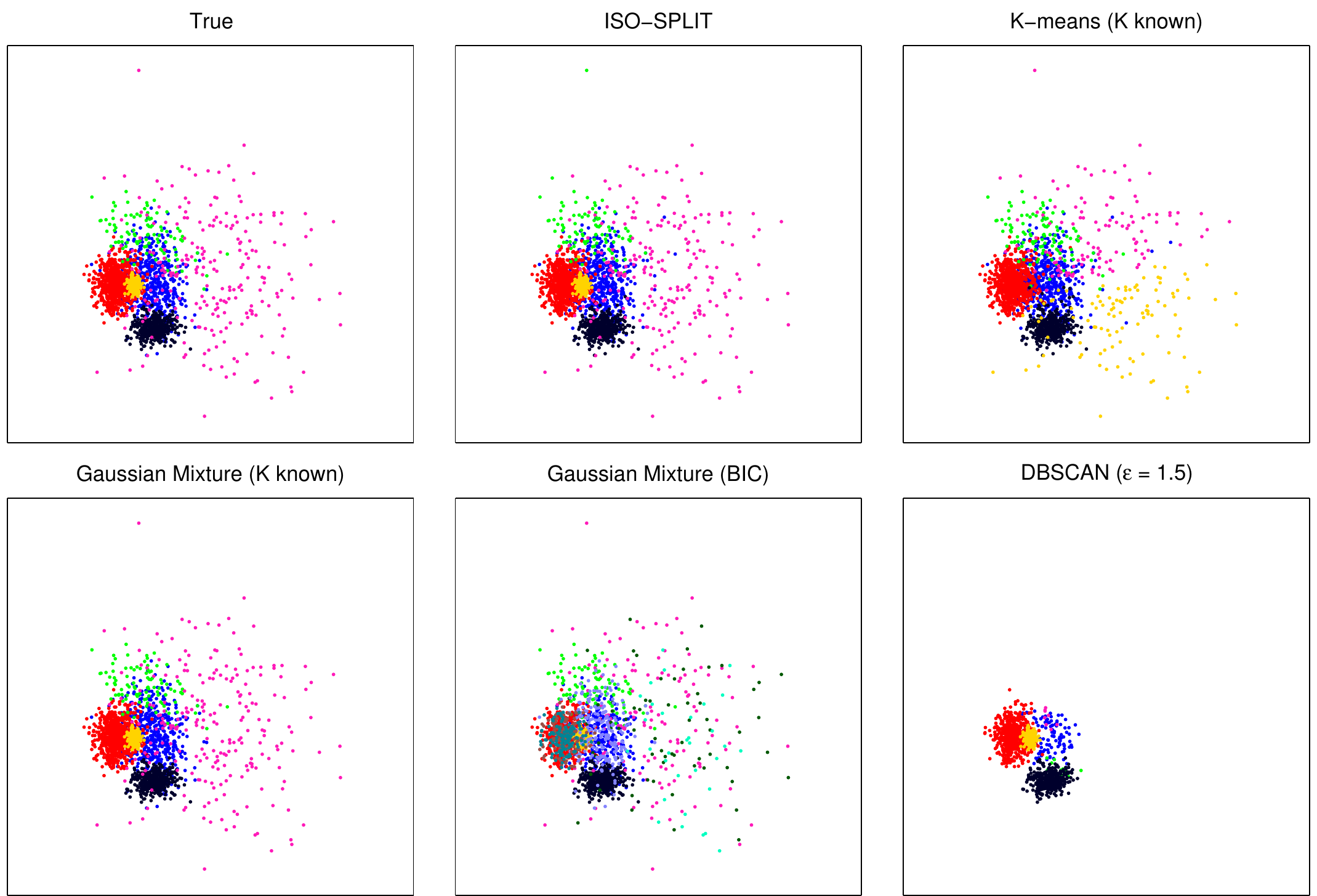}
\end{center}
\caption{
Projection onto two dimensions of an example run from \textbf{Simulation 5} with the number of dimensions equal to $p=6$. ISO-SPLIT performs well in this case whereas all other approaches struggle apart from GMM when given the correct $K$.
}
\label{fig:simulation5}
\end{figure}

\newcommand{\multicell}[2][c]{%
  \begin{tabular}[#1]{@{}l@{}}#2\end{tabular}}

\spacingset{1.0} 
\begin{table}
  \centering
\begin{tabular}{l|c|c|c|}
	 & \textbf{3 Clusters} & \textbf{6 Clusters} & \textbf{12 Clusters} \\
	\multicell{\textbf{Simulation 1 (Isotropic)}\\Gaussian, Isotropic, Equal pops.}  & & & \\ 
	\hline
  ISO-SPLIT & $98.7 \pm 0.1\%$ & $98.2 \pm 0.1\%$ & $96.6 \pm 0.7\%$ \\
  K-means (K known) & $99.0 \pm 0.1\%$ & $98.7 \pm 0.1\%$ & $98.4 \pm 0.0\%$ \\
  Gaussian Mixture (K known) & $98.9 \pm 0.1\%$ & $98.6 \pm 0.1\%$ & $97.9 \pm 0.4\%$ \\
  Gaussian Mixture (BIC) & $98.9 \pm 0.1\%$ & $98.6 \pm 0.1\%$ & $98.3 \pm 0.1\%$ \\
  DBSCAN ($\epsilon = 0.27$) & $94.8 \pm 0.4\%$ & $95.0 \pm 0.2\%$ & $94.3 \pm 0.2\%$ \\
  & & & \\
  \multicell{\textbf{Simulation 2 (Anisotropic)}\\Gaussian, \textbf{Anisotropic, Unequal pops.}}  & & & \\  
  \hline
  ISO-SPLIT & $94.7 \pm 2.2\%$ & $93.6 \pm 1.9\%$ & $94.1 \pm 1.3\%$ \\
  K-means (K known) & $85.3 \pm 3.7\%$ & $85.7 \pm 2.3\%$ & $80.5 \pm 1.6\%$ \\
  Gaussian Mixture (K known) & $98.4 \pm 0.2\%$ & $95.7 \pm 1.6\%$ & $91.2 \pm 1.5\%$ \\
  Gaussian Mixture (BIC) & $92.8 \pm 2.0\%$ & $95.4 \pm 1.0\%$ & $92.5 \pm 1.1\%$ \\
  DBSCAN ($\epsilon = 0.3$) & $66.7 \pm 3.3\%$ & $74.0 \pm 3.2\%$ & $66.8 \pm 2.9\%$ \\
  & & & \\
  \multicell{\textbf{Simulation 3 (Skewed)}\\\textbf{Non-Gaussian}, Anisotropic, Unequal pops.}  & & & \\ 
  \hline
  ISO-SPLIT & $92.2 \pm 2.3\%$ & $94.4 \pm 0.9\%$ & $86.5 \pm 3.8\%$ \\
  K-means (K known) & $85.8 \pm 3.2\%$ & $81.9 \pm 2.6\%$ & $79.5 \pm 1.5\%$ \\
  Gaussian Mixture (K known) & $78.0 \pm 2.0\%$ & $81.5 \pm 1.5\%$ & $79.4 \pm 1.3\%$ \\
  Gaussian Mixture (BIC) & $79.6 \pm 1.3\%$ & $83.6 \pm 1.2\%$ & $82.2 \pm 1.2\%$ \\
  DBSCAN ($\epsilon = 0.3$) & $78.5 \pm 4.1\%$ & $72.9 \pm 3.4\%$ & $74.1 \pm 2.7\%$ \\
  & & & \\
  \multicell{\textbf{Simulation 4 (Packed)}\\Gaussian, Isotropic, Equal pops., \textbf{Tightly packed}} & & & \\
  \hline
  ISO-SPLIT & $79.1 \pm 4.8\%$ & $55.3 \pm 5.5\%$ & $29.4 \pm 3.9\%$ \\
  K-means (K known) & $93.4 \pm 0.2\%$ & $91.5 \pm 0.2\%$ & $89.9 \pm 0.2\%$ \\
  Gaussian Mixture (K known) & $93.0 \pm 0.2\%$ & $91.0 \pm 0.3\%$ & $88.0 \pm 0.5\%$ \\
  Gaussian Mixture (BIC) & $92.9 \pm 0.2\%$ & $91.0 \pm 0.3\%$ & $86.4 \pm 1.1\%$ \\
  DBSCAN ($\epsilon = 0.3$) & $33.3 \pm 0.0\%$ & $16.7 \pm 0.0\%$ & $8.3 \pm 0.0\%$ \\
  & & & \\
  \multicell{\textbf{Simulation 5 (High-dimensional)}\\Gaussian, Isotropic, Equal pops., \textbf{\# Dims = 6}}  & & & \\ 
  \hline
  ISO-SPLIT & $88.0 \pm 3.1\%$ & $96.3 \pm 0.2\%$ & $82.1 \pm 3.3\%$ \\
  K-means (K known) & $76.4 \pm 4.0\%$ & $81.4 \pm 2.5\%$ & $74.8 \pm 2.4\%$ \\
  Gaussian Mixture (K known) & $99.1 \pm 0.1\%$ & $98.7 \pm 0.1\%$ & $87.1 \pm 2.3\%$ \\
  Gaussian Mixture (BIC) & $72.5 \pm 1.5\%$ & $81.1 \pm 2.1\%$ & $85.9 \pm 1.5\%$ \\
  DBSCAN ($\epsilon = 1.5$) & $56.7 \pm 5.3\%$ & $54.2 \pm 4.2\%$ & $40.1 \pm 4.5\%$ \\
	\hline
\end{tabular}
\caption{
\label{table:simulations}
Average accuracies over $20$ trial simulations with standard errors. All data, except for in Simulation 5, were generated in two dimensions ($p=2$). \textbf{Simulation 1}: Sample sizes were $500$ per cluster, $\xi=0$, $\zeta=0$, and cluster separations of $z_0=2.5$ (see text for details). \textbf{Simulation 2}: Sample sizes were randomly chosen between $100$ and $1000$ with $\xi=1.2$, $\zeta=2$, and cluster separations of $z_0=2.5$. \textbf{Simulation 3}: Same as Simulation 2 except with non-Gaussian/skewed clusters (see text for details). \textbf{Simulation 4}: Same as Simulation 1 except with clusters chosen closer together using $z_0=1.7$. \textbf{Simulation 5}: Same as the Simulation 2 except with a higher number of dimensions $p=6$.
}
\end{table}
\spacingset{1.45} 

\begin{figure}
\begin{center}
\includegraphics[width=5.5in]{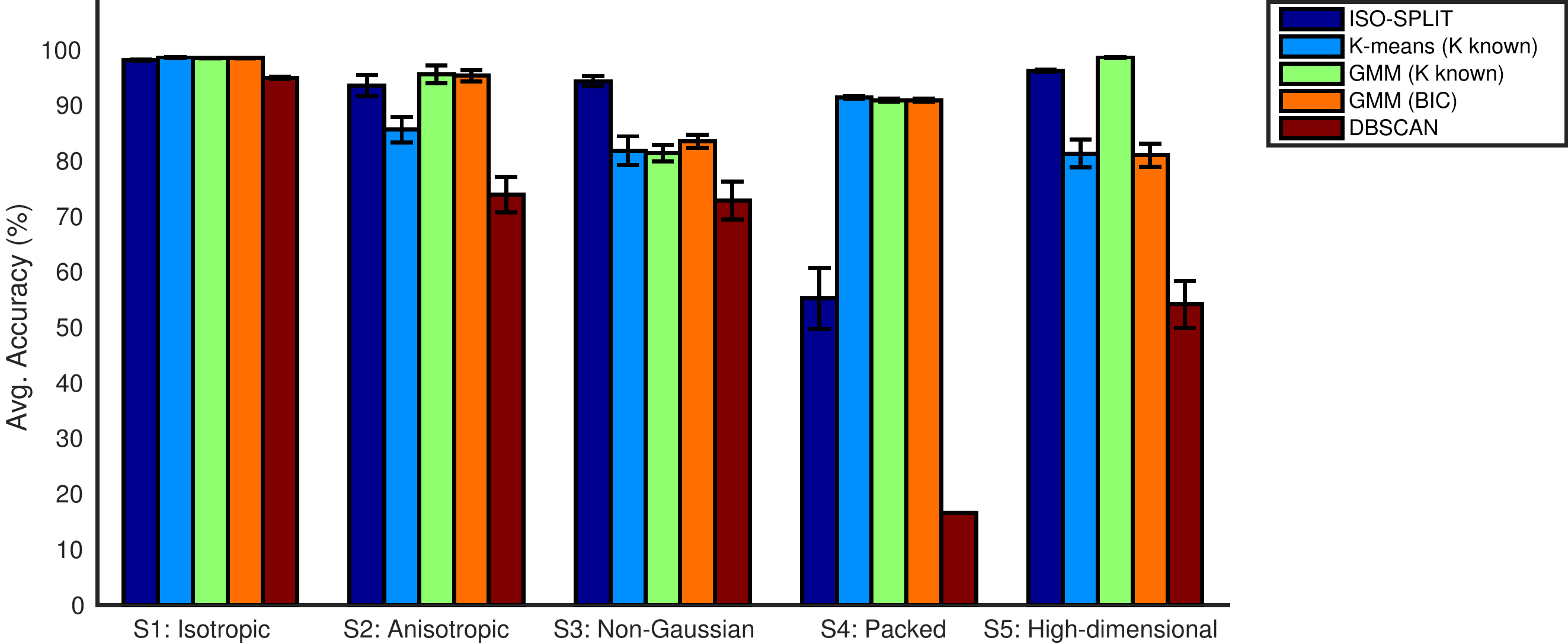}
\end{center}
\caption{
Results of simulations from Table \ref{table:simulations} with $6$ true clusters. Accuracies for ISO-SPLIT are at least comparable to the other approaches in all cases except for when clusters are tightly packed (Simulation 4). ISO-SPLIT yields the highest accuracy in the non-Gaussian clusters case (Simulation 3). ISO-SPLIT and GMM (BIC) were the only algorithms for which no adjustable parameters were specified.
}
\label{fig:result_bars}
\end{figure}

\section {Method comparison via simulation}
\label{algorithm_comparison}

A series of experiments were performed to compare the various approaches considered in this paper.
We use an accuracy measure
$$a=\frac{1}{K}\sum_{c=1}^K \min\left(\frac{n_{c,\pi(c)}}{n_c},\frac{n_{c,\pi(c)}}{n'_{\pi(c)}}\right),$$
where $K$ is the true number of clusters, $n_{i,j}$ is the number of points in true class $i$ labeled by the algorithm as class $j$, $n_i:=\sum_j n_{i,j}$ is the true size (population) of cluster $i$, and $n'_j:=\sum_i n_{i,j}$ the size of the found cluster $j$. Finally, $\pi(c)$ is the class number in the second labeling matching most closely to class $c$ in the true labeling (i.e., maximizing $n_{c,\pi(c)}$); note that this need not be a permutation of the original labels.
The summand of the measure $a$ is the smaller of {\em precision} and {\em recall}, and is thus a lower bound on
the F-measure \cite[Ch.~17]{zaki-book} (the latter being the harmonic mean of precision and recall).
We prefer this measure of cluster similarity over many others in the literature because it
weights each cluster equally, allowing us to assess fairly the performance when there is a wide range of cluster populations. (Such a metric is also the relevant one for the application presented in Sec.~\ref{s:spike}.)
Note also that the contribution of each cluster is the minimum of (a) the extent to which the cluster is not split, and (b) the extent to which the cluster is not merged with another cluster. Thus a contribution of $100\%$ means that the cluster is labeled perfectly with respect to both sensitivity (recall) and specificity. Unlike the F-measure, a value of $50\%$ could mean that the cluster was either split evenly or merged with a cluster of the same size.

Simulations in two dimensions and higher were performed by generating random samplings from a mixed multivariate Gaussian distribution (except for the Skewed simulation) with clusters corresponding to the individual Gaussian sub-populations. The centers, spreads, orientations, anisotropies, and populations of the Gaussians were varied randomly. Specifically, the random covariance matrix for each cluster was defined as
\[
\Sigma=R
\left(
\begin{array}{ccc}
e^{r_0\zeta + r_1\xi} & & 0 \\
 & \ddots &  \\
 0 &  & e^{r_0\zeta + r_n\xi} \\
\end{array}
\right)R^T,
\]
where $r_0,r_1,\dots,r_n$ are random numbers uniformly selected from $[-1,1]$, $\zeta$ is the spread variation factor, $\xi$ is the anisotropy variation factor and $R$ is a random rotation matrix. The cluster locations were chosen such that the clusters were packed tightly with the constraint that the solid ellipses corresponding to Mahalanobis distance $z\leq z_0$ did not intersect (see Appendix \ref{appendixPacking} for details). In Simulation 3 the Gaussian distributions were replaced by non-Gaussian distributions that were skewed in both dimensions. Specifically, the data points (prior to adjustment for the covariance matrix) were generated as $\tilde{R}[F(z_1),F(z_2)]$ for $z_j$ normally distributed with
\begin{equation}
\label{eq:nongaussian}F(z)=\log(|z+3|),
\end{equation}
and $\tilde{R}$ a random rotation matrix (fixed for each cluster). The histogram for the 1D non-Gaussian distribution is shown in Figure \ref{fig:nongaussian_histogram}, and examples of two-dimensional clusters are shown in Figure \ref{fig:simulation3}.

All experiments were performed on a Linux laptop with a 2.8GHz quad-core
i7 processor and 8GB RAM.
ISO-SPLIT was implemented in MATLAB with kernel routines in C++
(a URL for our software is given in Sec.~\ref{s:conc}).
The other clustering methods were implemented as follows.
We used a MATLAB implementation%
\footnote{Specifically we used the code from \citep{kmeanspp_sorber} corrected to use
  the original ``$D^2$ weighting'' recommended in \citep{kmeanspp}.}
 of k-means++
 with $100$ trials/restarts. GMM was performed using \citet{vlfeat} in MATLAB using EM iterations with 20 restarts. Both k-means and GMM had the advantage of being provided with the true number of clusters as input. A second run of GMM was used to automatically select the number of clusters using the Bayesian information criterion (BIC) \citep{BIC}. We used a multi-core C++ implementation of DBSCAN \citep{dbscan_dbp}, with $\epsilon$ tuned by hand for each simulation to yield the highest average accuracy measure.

All methods performed well in the Isotropic simulation under the conditions of equal populations, equal variances, and isotropic clusters, the ideal scenario for k-means (Figure \ref{fig:simulation1}).
For the second simulation, ISO-SPLIT and GMM performed best. As expected, k-means did not do as well since the clusters were not isotropic and had unequal populations. The unequal cluster densities caused problems for DBSCAN. See Figure \ref{fig:simulation2}.

The number of dimensions was increased to $6$ in the fifth simulation. In this case ISO-SPLIT performed at least comparably with the other methods. DBSCAN particularly struggled in this higher dimensional case. When $K$ was not known, GMM also struggled as illustrated in Figure \ref{fig:simulation5}.

In general, we see from the overall summary in Figure~\ref{fig:result_bars} that ISO-SPLIT did as well or better than all other methods except in Simulation 4 for which clusters were tightly packed, presumably because the tighter clusters were no longer separated by regions of significantly lower density (see Figure \ref{fig:simulation4}).

\section {Application to spike sorting data}
\label{s:spike}

The algorithms considered in this paper were applied to spike sorting of neural electrophysiological signals, that is, the clustering of spiking events in a time series into $K$ clusters that can often be associated with individual neurons. Clustering is a key step in spike sorting. More specifically, we took 7 channels of data from a set of adjacent electrodes in a multielectrode array which records voltages from an {\em ex vivo} monkey retina \citep{litke}; the 2 minutes of 20 kHz sampled time series data was supplied to us by the Chichilnisky Lab at Stanford. A set of points in $\mathbb{R}^{10}$ that needed to be clustered was created as follows. Firstly the time series was high-pass filtered at 300 Hz (which insures that the signal mean is zero), then windows of time series of length 60 samples were extracted in which the minimum voltage dropped below $-120 \mu$V. (We also applied an automatic procedure to remove windows that did not appear to be single spike events.) This gave $n=7275$ windows. The data in each window was upsampled by a factor of 3 using Hann-windowed sync interpolation, then negative-peak aligned to the central time point, then stacked to give a length-1029 column vector. Dimension reduction to $p=10$ dimensions was finally done using PCA; that is, the 1029-by-7275 matrix $A$ was replaced by the first 10 columns of $V^T A$, where $V$ is the matrix 
whose columns are the eigenvectors of $AA^T$ ordered by descending eigenvalue.
This dimension reduction accounted for a fraction $68\%$ of the total variance in the original data.
Columns of the resulting $A$ gave the 7275 points to cluster.

\begin{figure}
\begin{center}
\includegraphics[width=5.5in]{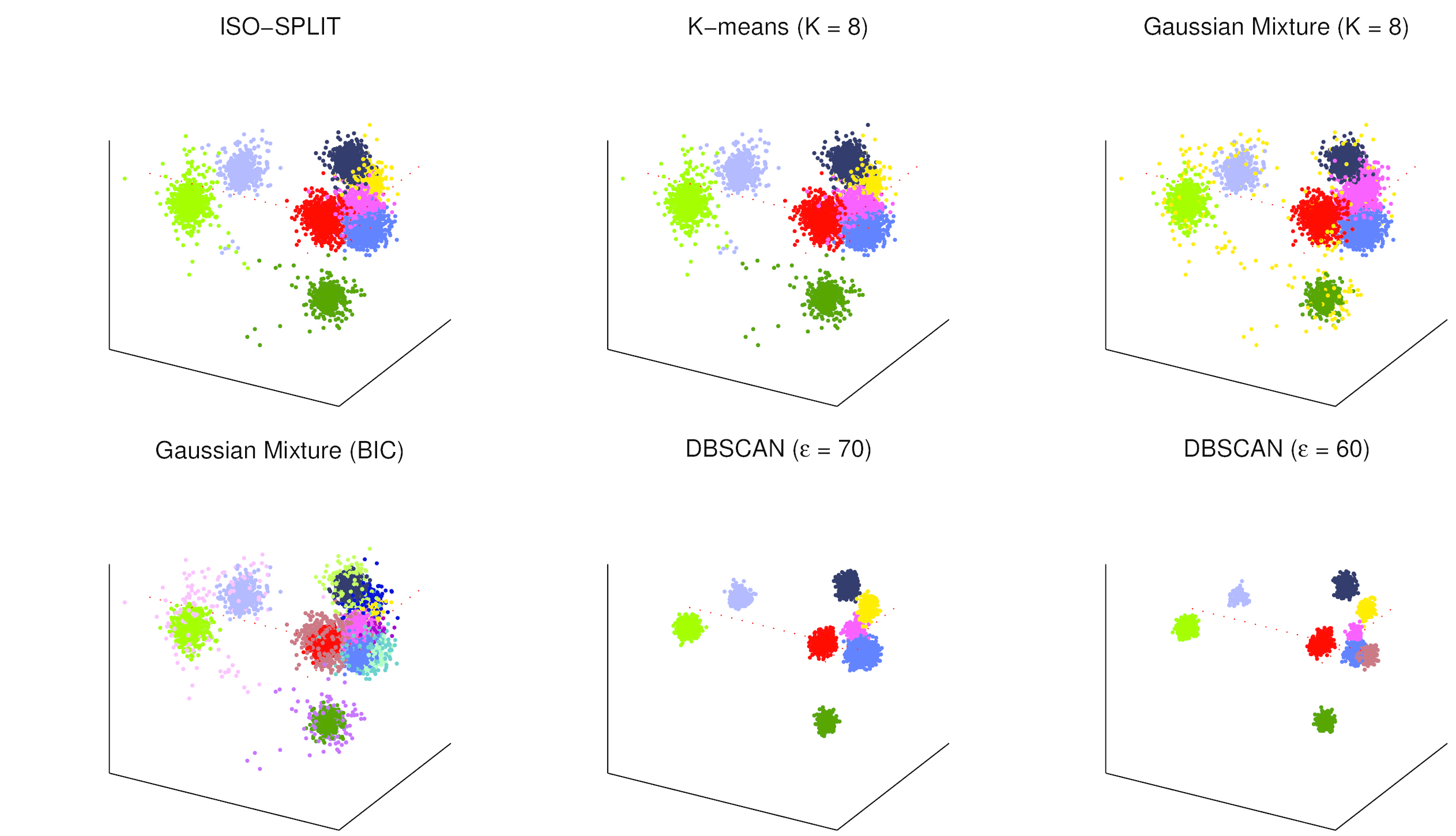}
\end{center}
\caption{
  Spike sorting clustering application on real data with $n=7275$ points in 10 dimensions. The first three dimensions of the 10-dimensional PCA feature space are shown, with coloring indicating cluster (neuronal unit) membership produced by each clustering algorithm indicated. The classifications were highly consistent between the various methods when k-means, GMM, and DBSCAN were tuned to yield the same number of clusters as ISO-SPLIT.
  The red dotted lines indicate axes intersecting at the origin.
}
\label{fig:real_data_1}
\end{figure}

\begin{figure}
\begin{center}
\includegraphics[width=5.5in]{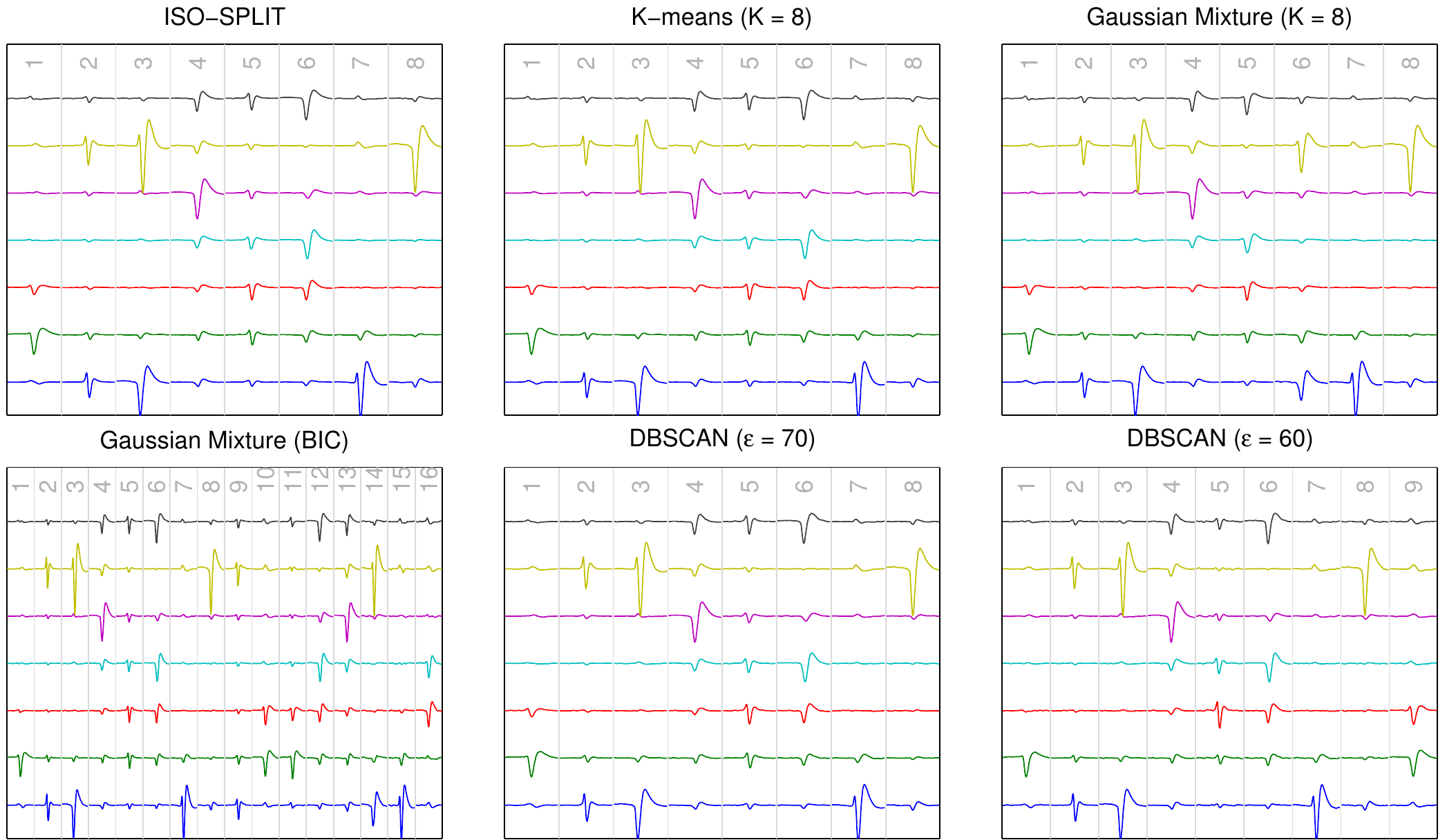}
\end{center}
\caption{
Spike waveforms corresponding to the centroids of the clusters from Figure \ref{fig:real_data_1}. Rows represent electrode channels and columns represent distinct neurons/clusters.
}
\label{fig:real_data_2}
\end{figure}

The resulting clustering (neuron labelings) are shown in Figure \ref{fig:real_data_1} with the centroid waveforms for each cluster displayed in Figure \ref{fig:real_data_2}. The adjustable parameters for k-means, GMM, and DBSCAN were carefully chosen to yield the same number of clusters ($K=8$) as ISO-SPLIT. Each of the algorithms produced a splitting into into qualitatively distinct centroid waveforms, suggesting that at least 8 detectable neural units are present. Various validation methods, and human judgements, are now commonly applied to determine whether each cluster is a single neural unit \citep[e.g.][]{Hill2011,validspike}; we do not attempt to perform that here.

We emphasize that ISO-SPLIT required no adjustment of free parameters, in contrast to all but the (poorly-performing) GMM with BIC. Indeed, when the BIC was used to select $K$ in GMM, many more clusters were identified, most of them duplicates, or artificially split clusters (see Figure \ref{fig:real_data_2}). An additional run of DBSCAN with $\epsilon$ reduced from $70$ to $60$ identified an additional cluster, demonstrating the sensitivity of the output to adjustments in this parameter. Further analysis is required to determine which results are closer to the truth.

\spacingset{1.0} 
\begin{table}
  \centering
  \begin{tabular}{l|c|c|c|}
	 & \textbf{3 Clusters} & \textbf{6 Clusters} & \textbf{12 Clusters} \\
	\multicell{\textbf{Simulation 1 (Isotropic)}\\Gaussian, Isotropic, Equal pops.}  & & & \\ 
	\hline
	ISO-SPLIT & $0.06$ & $0.12$ & $0.41$ \\
  K-means (K known) & $0.13$ & $0.57$ & $2.50$ \\
  Gaussian Mixture (K known) & $0.03$ & $0.16$ & $1.28$ \\
  Gaussian Mixture (BIC) & $0.15$ & $0.92$ & $9.78$ \\
  DBSCAN ($\epsilon = 0.27$) & $0.04$ & $0.04$ & $0.05$ \\
	& & & \\
	\multicell{\textbf{Simulation 2 (Anisotropic)}\\Gaussian, \textbf{Anisotropic, Unequal pops.}}  & & & \\  
	\hline
	ISO-SPLIT & $0.03$ & $0.09$ & $0.33$ \\
  K-means (K known) & $0.18$ & $0.61$ & $2.35$ \\
  Gaussian Mixture (K known) & $0.04$ & $0.14$ & $0.79$ \\
  Gaussian Mixture (BIC) & $0.18$ & $0.88$ & $6.63$ \\
  DBSCAN ($\epsilon = 0.3$) & $0.04$ & $0.04$ & $0.06$ \\
	& & & \\
	\multicell{\textbf{Simulation 3 (Skewed)}\\\textbf{Non-Gaussian}, Anisotropic, Unequal pops.}  & & & \\ 
	\hline
	ISO-SPLIT & $0.04$ & $0.08$ & $0.58$ \\
  K-means (K known) & $0.18$ & $0.54$ & $2.36$ \\
  Gaussian Mixture (K known) & $0.05$ & $0.13$ & $0.88$ \\
  Gaussian Mixture (BIC) & $0.36$ & $1.37$ & $9.79$ \\
  DBSCAN ($\epsilon = 0.3$) & $0.04$ & $0.04$ & $0.06$ \\
	& & & \\
	\multicell{\textbf{Simulation 4 (Packed)}\\Gaussian, Isotropic, Equal pops., \textbf{Tightly packed}} & & & \\
	\hline
	ISO-SPLIT & $0.04$ & $0.15$ & $0.63$ \\
  K-means (K known) & $0.17$ & $0.70$ & $3.08$ \\
  Gaussian Mixture (K known) & $0.06$ & $0.23$ & $1.28$ \\
  Gaussian Mixture (BIC) & $0.19$ & $1.07$ & $8.77$ \\
  DBSCAN ($\epsilon = 0.3$) & $0.04$ & $0.04$ & $0.05$ \\
	& & & \\
	\multicell{\textbf{Simulation 5 (High-dimensional)}\\Gaussian, Isotropic, Equal pops., \textbf{\# Dims = 6}}  & & & \\ 
	\hline
	ISO-SPLIT & $0.04$ & $0.17$ & $6.53$ \\
  K-means (K known) & $0.27$ & $0.86$ & $5.09$ \\
  Gaussian Mixture (K known) & $0.02$ & $0.09$ & $0.68$ \\
  Gaussian Mixture (BIC) & $0.30$ & $1.21$ & $7.51$ \\
  DBSCAN ($\epsilon = 1.5$) & $0.04$ & $0.05$ & $0.09$ \\
\end{tabular}
\caption{
\label{table:simulations2}
Average computation times in seconds per trial run for the simulations of Table \ref{table:simulations}.
}
\end{table}
\spacingset{1.45} 

\section {Computational efficiency}

Each iteration of ISO-SPLIT comprises two steps. The first step, projection onto 1D space, has computation time $O(n_0 p)$ where $p$ is the number of dimensions and $n_0<n$ is the number of points involved in the two clusters of interest. The second step is 1D clustering using the Hartigan test and isotonic regression and has time complexity $O(n_0)$. Due to the complexity of the cluster redistributions at each step, it is difficult to theoretically estimate the number iterations required for convergence.

Table \ref{table:simulations2} shows empirical average computation times for the simulations of Table \ref{table:simulations}. Overall prefactors in the running times should not be given too much meaning, since they are highly implementation-dependent. In almost every case, GMM with unknown $K$ took the longest since the algorithm needed to be run several times to find the optimal number of clusters using the BIC. Even when $K$ was known, GMM and k-means were rerun many times (20 for GMM and 100 for k-means), and therefore took longer on average than ISO-SPLIT in almost every case. With a highly optimized C++ implementation \citep{dbscan_dbp}, DBSCAN was the most efficient algorithm of those considered. There was a significant increase in ISO-SPLIT's run time from 6 to 12 clusters (especially in Simulation 5), but it should be noted that both $K$ and $n$ were increased by a factor of $2$, since the number of points per cluster was fixed.

\begin{figure}
\begin{center}
\includegraphics[width=6in]{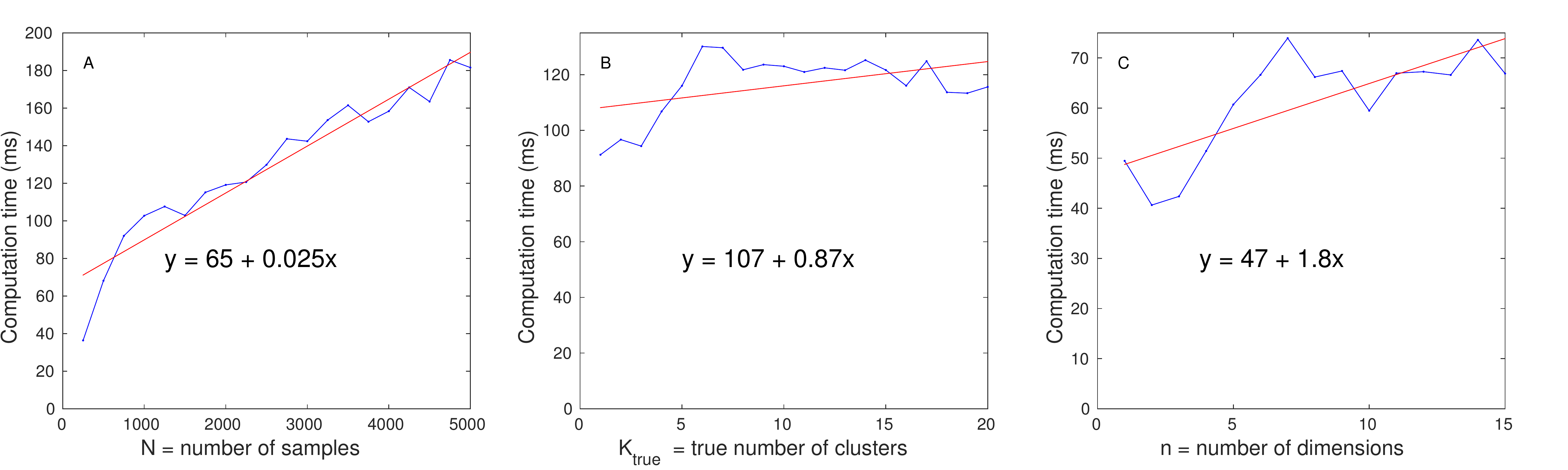}
\end{center}
\caption{
Empirical dependence of average computation time on three simulation parameters, averaged over $20$ repeats; (A) number of points per sample, with number of true clusters fixed at $6$ and $p$ fixed at $2$; (B) number of true clusters, with $n$ fixed at $1000$ and $p$ fixed at $2$; (C) number of dimensions with $n$ fixed at $1000$ and number of true clusters fixed at $6$, with noise dimensions added past the first two dimensions.
}
\label{fig:computation_times_01}
\end{figure}

Future work will investigate the theoretical bounds on the number of iterations required for convergence. Here we present empirical estimates for the scaling properties with increasing values of $n$ (the number of samples), $p$ (the number of dimensions), and $K_\text{true}$ (the true number of clusters). The most time-consuming step in each iteration (isotonic regression) is independent of both $p$ and $K_\text{true}$ but the unknown quantity is the number of iterations required for convergence. Figure \ref{fig:computation_times_01} shows the results of three simulations suggesting that computation time scales linearly with all three simulation parameters.

Using code profiling in MATLAB, we found that the majority of time (approx. 90\%) was spent performing 1D clustering (Section~\ref{clustering_1d}) as compared with finding the best pair of clusters to compare, computing centroids, and projecting data onto lines. Of course this figure will depend on $p$, $K$, $n$, and the structure of the data.

\section {Discussion}

We have shown that, for the target application of spike sorting, our new technique produces results that are consistent with those of standard clustering techniques (k-means, GMM, DBSCAN). Yet the key advantage of ISO-SPLIT is that it does not require selection of scale parameters nor the number of clusters. This is very important in situations where manual processing steps are to be avoided.
Automation is also critical when hundreds of clustering runs must be executed within a single analysis, e.g., applications of spike sorting with large electrode arrays. Furthermore, the accuracy of ISO-SPLIT appears to exceed that of standard techniques in the context of many simulations performed in this study. Most notably, it excels when clusters are non-Gaussian with varying populations, orientations, spreads, and anisotropies.

While ISO-SPLIT outperforms standard methods in situations satisfying the assumptions of the method, the algorithm has general limitations and is not suited for all contexts. Because ISO-SPLIT depends on statistically significant density dips between clusters, erroneous merging occurs when clusters are positioned close to one another (see the Packed simulation). Certainly this is a challenging scenario for all algorithms, but k-means or mixture models are better suited to handle these cases. On the other hand,
if the underlying density has dips which separate clusters, ISO-SPLIT will
find them for sufficiently large $n$.

Since the one-dimensional tests are repeated at every iteration as the clusters are merged and redistributed, it is not possible to interpret the significance threshold of the one-dimensional tests in the context of the multi-dimensional clustering. Further exploration is therefore needed to understand the expected false splitting and merging rates.

Our theory depends on the assumption that the data arise from a continuous probability distribution. While no particular noise model is assumed, we do assume that, after projection onto any 1D space, the distribution is locally well approximated by a uniform distribution. This condition is satisfied for any smooth probability distribution. In particular, it guarantees that no two samples have exactly the same value (which could lead to an infinite estimate of pointwise density). Situations where values are drawn from a discrete grid (e.g., an integer lattice) will fail to have this crucial property. One remedy for such scenarios could be to add random offsets to the datapoints to form a continuous distribution.

Clusters with non-convex shapes may be well separated in density but not separated by a hyperplane (Figure \ref{fig:example_dbscan}). In these situations, alternative methods such as DBSCAN are preferable. But even when clusters are convex, a pair may be oriented such that the separating hyperplane is not orthogonal to the line connecting the centroids.

While each iteration is efficient (essentially linear in a subset of the number of points of interest), computation time may be a concern since the number of iterations required to converge is unknown. Empirically, total computation time appears to increase linearly with the number of clusters, the number of dimensions, and the sample size.

As mentioned above, a principal advantage of ISO-SPLIT is that it does not require parameter adjustments. Indeed, the core computational step is isotonic regression, which does not rely on any tunable parameters. Two parameters are fixed once and for all, the threshold of rejecting the unimodality hypothesis for the 1D tests, and $K_\text{initial}$, the initial number of clusters. In Appendix \ref{appendixSensitivity} we argue that the algorithm is not sensitive to these values over reasonable ranges.

\section{Conclusion}
\label{s:conc}

A multi-dimensional clustering technique, ISO-SPLIT, based on density clustering of one-dimensional projections was introduced. The algorithm was motivated by the electrophysiological spike sorting application. Unlike many existing techniques, the new algorithm does not depend on adjustable parameters such as scale or \emph{a priori} knowledge of the number of clusters. Using simulations, ISO-SPLIT was compared with k-means, Gaussian mixture, and DBSCAN, and was shown to outperform these methods in situations where clusters were separated by regions of relatively lower density and where each pair of clusters could be largely split by a hyperplane. ISO-SPLIT was especially effective for non-Gaussian cluster distributions. Future research will focus on applying the algorithm to additional real-world problems as well as improving computational efficiency.

A MATLAB/C++ implementation of ISO-SPLIT is freely available
at the following URL:\\
{\tt http://github.com/magland/isosplit}

\section*{Acknowledgments}

We have benefited from useful discussions with Leslie Greengard,
Marina Spivak, Bin Yu, and Cheng Li, and from the comments of the anonymous reviewers.
We are grateful for EJ Chichilnisky and his research group
for supplying us with the retinal neural recording data used
in section~\ref{s:spike}.

\appendix 

\algrenewcomment[1]{\(\triangleright\) #1}

\spacingset{0.95} 
\begin{algorithm}
\caption{}
\begin{algorithmic}
\Function{UpDownIsotonic}{x}
	\State \Comment{Isotonic regression for increasing followed by decreasing}
	\State $n \gets \text{length}(x)$
	\State $b \gets \text{FindOptimalB}(x)$
	\State $y^{(1)} \gets \text{IsotonicIncreasing}([x_1,\dots,x_b])$
	\State $y^{(2)} \gets \text{IsotonicDecreasing}([x_b,\dots,x_n])$
	\State $y \gets [y^{(1)}_1,\dots,y^{(1)}_b,y^{(2)}_2,\dots,y^{(2)}_{n-b+1}]$
	\State \Return $y$
\EndFunction
\Statex
\Function{FindOptimalB}{x}
	\Statex \Comment{Find where to switch direction}
	\State $x^{(1)} \gets x$
	\State $x^{(2)} \gets -\text{Reverse}(x)$
	\State $\mu^{(1)} \gets \text{PAVA-MSE}(x_1)$	
	\State $\mu^{(2)} \gets \text{Reverse}(\text{PAVA-MSE}(x^{(2)}))$
	\State Find $b$ to minimize $\mu^{(1)}_b+\mu^{(2)}_b$
	\State \Return $b$
\EndFunction
\Statex
\Function{Reverse}{$[x_1,\dots,x_n]$}
	\State \Comment{Reverse the ordering}
	\State \Return $[x_n,x_{n-1},\dots,x_1]$
\EndFunction
\Statex
\Function{PAVA-MSE}{$[x_1,\dots,x_n]$,$[w_1,\dots,w_n]$}          
	\State \Comment{Modified PAVA to return MSE at every index}
	\State $i \gets 1$, $j \gets 1$
	\State $\text{count}[i] \gets 1$, $\text{wcount}[i] \gets w_j$
	\State $\text{sum}[i] \gets w_j x_j$, $\text{sumsqr}[i] \gets w_j x_j^2$
	\State $\mu_j \gets 0$
	\State
	\For{$j=2\dots n$}
		\State $i \gets i+1$
		\State $\text{count}[i] \gets 1$, $\text{wcount}[i] \gets w_j$
		\State $\text{sum}[i] \gets w_j x_j$, $\text{sumsqr}[i] \gets w_j x_j^2$
		\State $\mu_j \gets \mu_{j-1}$
		\Loop
			\If{$i=1$}
				\textbf{ break} 
			\EndIf
			\If{$\text{sum}[i-1]/\text{count}[i-1]<\text{sum}[i]/\text{count}[i]$}
				\textbf{ break} 
			\Else \Comment{ Merge the blocks}
				\State $\mu_{\text{before}}\gets\text{sumsqr}[i-1]-\text{sum}[i-1]^2/\text{count}[i-1]$
				\State $\mu_{\text{before}}\gets \mu_{\text{before}}+\text{sumsqr}[i]-\text{sum}[i]^2/\text{count}[i]$
				\State $\text{count}[i-1] \gets \text{count}[i-1]+\text{count}[i]$, $\text{wcount}[i-1] \gets \text{wcount}[i-1]+\text{wcount}[i]$
				\State $\text{sum}[i-1] \gets \text{sum}[i-1]+\text{sum}[i]$, $\text{sumsqr}[i-1] \gets \text{sumsqr}[i-1]+\text{sumsqr}[i]$
				\State $\mu_{\text{after}}\gets\text{sumsqr}[i-1]-\text{sum}[i-1]^2/\text{count}[i-1]$
				\State $\mu_j\gets \mu_j+\mu_{\text{after}}-\mu_{\text{before}}$
				\State $i\gets i-1$
			\EndIf
		\EndLoop
	\EndFor
	\State \Return $\mu$
\EndFunction

\end{algorithmic}
\label{alg:PAVA2}
\end{algorithm}
\spacingset{1.45} 

\section {Up-down isotonic regression}
\label{appendixUpdown}

In this section we outline a computationally efficient variant of isotonic regression that provides the critical step in the kernel operation of ISO-SPLIT. Isotonic regression is a non-parametric method for fitting an ordered set of real numbers by a monotonically increasing (or decreasing) function. Suppose we want to find the best least-squares approximation of the sequence $x_1,\dots,x_n$ by a monotonically increasing sequence. Considering the more general problem that includes weights, we want to minimize the objective function
\begin{equation}
F(y)=\sum_{i=1}^n w_i(y_i-x_i)^2,
\label{eq:least_squares}
\end{equation}
subject to
$$y_1\leq y_2\leq\dots\leq y_n.$$
This may be solved in linear time using the pool adjacent violators algorithm (PAVA) \citep{pava}; we do not include the full pseudocode for this standard algorithm but note that it is essentially the same as PAVA-MSE in Algorithm~\ref{alg:PAVA2}.

As discussed above, ISO-SPLIT depends on a variant of isotonic regression, which we call \emph{updown isotonic regression}. In this case we need to find a turning point $y_b$ such that $y_1\leq y_2\leq\dots\leq y_b$ and $y_b\geq y_{b+1}\dots\geq y_n$. Again we want to minimize $F(y)$ of Equation~\eqref{eq:least_squares}. One way to solve this is to use an exhaustive search for $b\in\{1,\dots,n\}$. However, this would have $O(n^2)$ time complexity.

A modified PAVA that finds the optimal $b$ for the updown case in linear time is presented in Algorithm~\ref{alg:PAVA2}. The idea is to perform isotonic regression from left to right and then right to left using a modified algorithm where the mean-squared error is recorded at each step. The turning point is then chosen to minimize the sum of the two errors.

Downup isotonic regression is also needed by the algorithm. This procedure is a straightforward modification of updown in Algorithm~\ref{alg:PAVA2} by negating both the input and output.

\section {Sensitivity to parameters}
\label{appendixSensitivity}

In this work we have claimed that ISO-SPLIT does not require parameter adjustments which depend on the application or nature of the dataset, thus facilitating fully automated clustering. However, practically speaking, there are two values mentioned in this paper that need to be set. In this section we demonstrate that the algorithm is not sensitive to these choices provided that they fall within a reasonable range.

First, the threshold $\tau_n=\alpha/\sqrt{n}$ for rejecting the unimodality hypothesis needs to be chosen. To demonstrate, we ran Simulation 2 again with varying $\alpha$. The results are found in Table \ref{table:alpha_dependence}. For $\alpha$ between $1.2$ and $2.0$, the accuracies remained virtually constant.

  \begin{table}[t]
\centering
    \begin{tabular}{c|c|c|}
	\textbf{$\alpha$} & \textbf{Accuracy} & \textbf{Time (s)} \\
	\hline
0.90 & 63\% & 0.15 \\
1.00 & 67\% & 0.14 \\
1.10 & 86\% & 0.17 \\
1.20 & 89\% & 0.18 \\
1.30 & 91\% & 0.14 \\
1.40 & 89\% & 0.12 \\
1.50 & 89\% & 0.12 \\
1.60 & 91\% & 0.13 \\
1.70 & 87\% & 0.13 \\
1.80 & 88\% & 0.12 \\
1.90 & 88\% & 0.13 \\
2.00 & 90\% & 0.14 \\
2.10 & 85\% & 0.11 \\
2.20 & 70\% & 0.12 \\
2.30 & 64\% & 0.10 \\
\end{tabular}
\caption{
\label{table:alpha_dependence}
Results of Simulation 2 (ISO-SPLIT, 6 clusters) repeated with varying threshold parameter $\alpha$ demonstrating insensitivity to the choice of this parameter (when within a reasonable range).
}
\end{table}

The second value to be set is $K_{\text{initial}}$, the number of clusters used to initialize the algorithm. Our hypothesis is that the choice of this parameter will have no affect on the output assuming that it is chosen large enough. This is supported in Table \ref{table:initial_K_dependence}. As discussed above, larger values of $K_{\text{initial}}$ will lead to significantly longer run times (quadratic time dependence), so it is important not to set this value to be much higher than necessary. 

\begin{table}[t]
  \centering
\begin{tabular}{c|c|c|}
	\textbf{Initial $K$} & \textbf{Accuracy} & \textbf{Time (s)} \\
	\hline
	3 & 44\% & 0.03 \\
6 & 87\% & 0.08 \\
12 & 92\% & 0.09 \\
24 & 90\% & 0.13 \\
48 & 89\% & 0.19 \\
96 & 88\% & 0.32 \\
\end{tabular}
\caption{
\label{table:initial_K_dependence}
Results of Simulation 2 (ISO-SPLIT, 6 clusters) repeated with varying $K_\text{initial}$ demonstrating insensitivity to the choice of this parameter (above a certain value). As expected, computation time increased for larger $K_\text{initial}$.
}
\end{table}

\section {Packing Gaussian clusters for simulations}
\label{appendixPacking}

Simulations 1-5 required automatic generation of synthetic datasets with fixed numbers of clusters of varying densities, populations, spreads, anisotropies, and orientations. The most challenging programming task was to determine the random locations of the cluster centers. If clusters were spaced out too much then the clustering would be trivial. On the other hand, overlapping clusters cannot be expected to be successfully separated. Here we briefly describe a procedure for choosing the locations such that clusters are tightly packed with the constraint that the solid ellipsoids corresponding to Mahalanobis distance $z_0$ do not intersect. Thus $z_0$ is a constant controlling the tightness of packing, fixed for each simulation. In above simulations we considered $z_0=2.5\text{ and }1.7$.

The clusters are positioned iteratively, one at a time. Each cluster is positioned at the origin and then moved out radially in small increments of a random direction until the non-intersection criteria is satisfied. Thus we only need to determine whether two clusters defined by $(\mu_1,\Sigma_1)$ and $(\mu_2,\Sigma_2)$ are spaced far enough apart. Here $\mu_j$ are the cluster centers and $\Sigma_j$ are the covariance matrices. The problem boils down to determining whether two arbitrary $p$-dimensional ellipsoids intersect. Surprisingly this is a nontrivial task, especially in higher dimensions, but an efficient iterative solution was discovered by \citet{ellipsoid-distance}. For the present study, the Lin--Han algorithm was implemented in MATLAB.

\bibliography{isosplit-revision1}
\end{document}